\documentstyle[aas2pp4]{article}

\newcommand{\beq}{\begin{equation}}
\newcommand{\eeq}{\end{equation}}
\def\beqa{\begin{eqnarray}}
\def\eeqa{\end{eqnarray}}
\newcommand{\lsim}{\lesssim}
\newcommand{\gsim}{\gtrsim}

\def\fl{Friedmann-Lema\^\i tre}
\def\flm{Friedmann-Lema\^\i tre model}
\input psfig

\begin{document}

\title{IS COSMOLOGY SOLVED?\\
An  Astrophysical Cosmologist's Viewpoint}

\author{P. J. E. Peebles}
\affil{Joseph Henry Laboratories, Princeton University,\\
and Princeton Institute for Advanced Study}

 
\begin{abstract}
We have fossil evidence from the thermal background radiation
that our universe expanded from a considerably hotter denser
state. We have a well defined, testable, and so far quite
successful theoretical description of the expansion: the 
relativistic \fl\ cosmological model. The observational successes of this model are impressive but I think hardly enough for a convincing scientific case. One way to see the limitations is to compare the lists of observational constraints and free hypotheses within the model; they have similar lengths. Another way to assess the state of the cosmological tests is to consider the search for concordant measures of the mass density parameter and the cosmological constant. The scorecard shows that the high density Einstein-de~Sitter model is seriously challenged, but that there is not much to choose between the grades for low mass density models with and without a cosmological constant. That is, it is hard to argue that the \flm\ is strongly overconstrained, the usual criterion for a mature theory. Work in progress will significantly improve the situation and may at last yield a compelling test.  
If so, and the \flm\ survives, it will close one line of research 
in cosmology: we will know the outlines of what happened as our
universe expanded and cooled from high density. It will not end research, of course; some of us will occupy ourselves with the
details of how galaxies and other large-scale structures came to
be the way they are, others with the issue of what our universe
was doing before it was expanding. The former is being driven by
rapid observational advances. The latter is being driven mainly by
theory, but there are hints of observational guidance. 
\end{abstract}


\section{Introduction}
Since few of us can see any indication we are nearing the end of
search and discovery in cosmology we have to adopt a convention for the meaning of ``cosmology solved.'' I take it to be a positive outcome of accurate and well cross-checked tests of the
relativistic \fl\ cosmological model. We are not there 
yet, but all signs are that now, some seven decades after the
first of the tests were proposed, we may be approaching a major closure. 

Many of my colleagues have concluded that the observational
successes of the \flm\ combined with its logical plausibility
already make the case, and it is time to move on to the  
issue of how initial conditions for this model were set by the
deeper physics of the very early universe, and how that led to
the origin of the world as we know it. This positive attitude is
healthy but maybe a little incomplete. I am taken by 
Willem de Sitter's (1931) remark: ``It should not be 
forgotten that all this talk about the universe involves a
tremendous extrapolation, which is a very dangerous operation.''
Observational advances since then have greatly reduced the
danger, but I think should leave us with a sense of wonder at
the successes in probing the large-scale nature of the physical
universe and caution in deciding just how well we understand the
situation.

A satisfactory understanding is easily defined: there must 
be more pieces of evidence than parameters we are free 
to adjust to fit the evidence. In \S 2 I comment on the still 
uncomfortably similar lengths of the lists of hypotheses 
and observational constraints in cosmology. Many cosmological tests constrain the two dimensionless parameters $\Omega _m$ and $\Omega _\Lambda$ that measure the relative contributions of matter and Einstein's cosmological constant to the expansion rate in the \flm . If we can establish that concordant values of these parameters follow from many more than two observational constraints we will have an important positive test of the model. I argue in \S 3 that there still is an uncomfortably large number of open issues: the parameters are not strongly overconstrained. In short, many commonly discussed elements of cosmology still are on dangerous ground. Work in progress promises to improve the situation; the community will be following the results with great interest to learn whether this aspect of cosmology may at last be declared ``solved.''

\section{Is Our Cosmology Predictive?}

\subsection{The Expanding Universe}

The first part of Table~1 refers to the idea that our universe has expanded from a considerably hotter denser state. Here, as Joe Silk describes in his contribution to these proceedings, we are on reasonably safe ground. Distant
objects, whose recession velocities approach the velocity of
light, are quite close to isotropic around us. Since distant
galaxies seem to be equally good homes for observers the
straightforward interpretation is that the universe is close to
homogeneous in the large scale average. In a homogeneously
expanding universe 
the recession velocity is proportional to the distance. This is
Hubble's law; it is observationally well established. We are
in a uniform sea of cosmic background radiation, the
CBR, with a spectrum that is quite close to thermal at $T=2.73$~K.
The only known explanation 
is relaxation to statistical equilibrium. This could not have
happened in the universe as it is now because space is
transparent: distant galaxies are observable as radio sources at
CBR wavelengths. The inference is that the CBR is a remnant from a
time when the universe was denser, hotter, and optically thick. That is, we have direct fossil evidence of the expansion and cooling of the universe.

All these results follow by symmetry arguments with conventional local physics; one does not need the full machinery of general relativity theory. Relativity is probed in more detailed tests. 
 
\subsection{The Cosmological Tests}
 
Joe Silk and Michael Turner discuss another likely fossil remnant of a time when our universe was very different from now. In the hot \flm\ helium and other light elements were produced in  thermonuclear reactions as
the universe expanded and cooled through temperatures $kT\sim
1$~MeV. This result is the first entry in the second part of Table~1. It uses the \fl\ expression for the expansion rate,
\beq
	{1\over 2}\left( da\over dt\right) ^2 - {GM\over a} =
			\hbox{constant},
	\label{eq:exprate}
\eeq
where 
\beq
  	M(t) = {4\pi\over 3}\rho a^3,
\eeq
and $a(t)$ is the expansion factor (such that the distance between conserved objects scales with the expansion of the universe as $a(t)$). The mass density $\rho (t)$ includes rest mass and the mass equivalent of energy; equation~(\ref{eq:exprate}) is a relativistic expression. But since you can guess at its form by analogy to Newtonian mechanics it is not a very deep application of general relativity theory. For that we must consider more of the cosmological tests.

The baryon density is an adjustable parameter in this theory. It is impressive that the value of $\Omega _{\rm baryon }$ that yields a satisfactory fit to the the observed abundances of the light element fits the astronomical surveys of the baryon density (Fukugita, Hogan, \&\ Peebles 1998), but the check is only good to a factor of three or so. The wanted baryon density is less than the mass density parameter $\Omega _m$ indicated by dynamical studies of the motions of galaxies relative to the general expansion (as discussed in \S 3.1). That is remedied by an hypothesis, that the mass of the universe is dominated by nonbaryonic matter. The straightforward reading of the dynamical estimates is that $\Omega _m$ is less than unity, contrary to the simple Einstein-de~Sitter case. A popular remedy is another postulate, that the mass of the universe is dominated by dark matter outside the concentrations of galaxies. 

We have a check on $\Omega _m$, from the magnificent work by two groups (Perlmutter {\it et al.} 1998a,b; Reiss {\it et al.} 1998) on the curvature of the redshift-magnitude ($z$-$m$) relation for supernovae of type Ia. A cautionary note is in order, however. The most distant supernovae are fainter than would be expected in the Einstein-de~Sitter case. How do we know that is not because the more distant supernovae are less luminous? The authors present careful checks, but the case has to be indirect: no one is going to examine any of these
supernovae up close, let alone make the trip back in time to compare distant supernovae to nearer examples. In short, the
supernovae measurement is a great advance, beautifully and
carefully done, but it does not come with a guarantee. The point is obvious to astronomers but not always to their colleagues in physics, and so might well be encoded in the Tantalus Principle: in astronomy you can look but never touch (with a few exceptions, such as objects in the Solar System, that are quite irrelevant for our purpose). 

The straightforward reading of the SNeIa $z$-$m$ relation within the \flm\ is that $\Omega _m$ is well below unity, consistent with dynamics, and that there is a significant contribution to the stress-energy tensor from Einstein's cosmological constant $\Lambda$ (or a term in the stress-energy tensor that acts like $\Lambda$). The latter has to be counted as another hypothesis, of course. This in turn can be checked by still more cosmological tests, such as the expansion time. But as discussed in the next section we don't yet have the wanted precision. 

Our tour of the second set of tests in Table~1, that probe  space-time geometry, shows no postulates that appear artificial, which is encouraging. But we do see that each constraint is met with a new free parameter, which is a dangerous operation, to quote de~Sitter (1931). 

\subsection{Structure Formation}

The third part of Table~1 refers to tests of the \flm\ from the condition that it admit a theory for the origin of cosmic structure: galaxies and all that. Here the danger is that we are testing two theories, cosmology and structure formation. Within the gravitational instability picture the latter requires a prescription for the important dynamical actors --- it may include cold dark matter, massive neutrinos, cosmic strings, or other fields --- and the character of the departures from homogeneity at high redshift. A commonly discussed model assumes Gaussian adiabatic fluctuations in cold dark matter, baryons, massless neutrinos, and the thermal cosmic background radiation (the CBR). That leaves one free function, the power spectrum of the initial mass density fluctuations, to fit to two functions, the spectra of fluctuations in the present space distribution of the mass and in the present angular distribution of the CBR. It is impressive that we can adjust the one function to match both sets of observations. But as discussed in \S 3.4 we do not yet have firm evidence that the initial conditions are Gaussian, or that they are adiabatic, or that the model takes account of all the important dynamical actors. We expect to have a check: if precision measurements in progress of the CBR and the large-scale matter distributions match in all detail the predictions of one of the simple models now under discussion, for reasonable values of the cosmological parameters, it will make believers of us all (or at least many of us). But before deciding to become a believer it might be wise to wait to see what the measurements reveal.

\section{A Scorecard}

I turn now to some details. The scorecard in Table~2 shows three parameter choices. In the Einstein-de~Sitter case the density parameter is $\Omega _m= 1$ and
space curvature and the cosmological constant $\Lambda$ are
negligibly small. The second case has zero
space curvature and positive $\Lambda$, the third
negligibly small $\Lambda$ and negative space curvature. 
A passing grade ($\surd$) means the
measurements are consistent with the parameters within the \flm,
a negative grade ({\bf X}) that there seems to be a significant 
inconsistency. The greater the number of question marks the greater the level of doubt about the grade.

\subsection{Dynamical Mass Estimates}

Here is an example that illustrates features common to many
dynamical estimates of the mean mass density. 

We are near the edge of a concentration of galaxies that de
Vaucouleurs (1956) called the Local Supercluster. It is centered
near the Virgo cluster, at distance 
\beq
	R = 12h^{-1}\hbox{ Mpc},
\eeq
where Hubble's constant is written as 
\beq
	H_o=100h\hbox{ km s}^{-1}\hbox{ Mpc}^{-1},
\label{eq:Ho}
\eeq
and the dimensionless parameter is thought to be in
the range $0.5\lsim h\lsim 0.8$. In the \flm\ the gravitational
attraction of the mass excess in this region 
produces a peculiar motion of inflow (relative to the general
expansion of the universe). In linear perturbation theory the
mass conservation law relates the peculiar velocity $\vec v(\vec
r,t)$ and the mass density contrast 
$\delta (\vec r,t)=\delta\rho /\rho$ by the equation 
\beq
	{\nabla\cdot\vec v\over a} = -{\partial\delta\over\partial t}
	= -{\dot D\over D}\delta.
\label{eq:mass_con}
\eeq
The divergence is with respect to comoving coordinates {\bf x},
where a physical length interval is
$\delta {\bf r}=a(t)\delta {\bf x}$ (and the expansion parameter $a(t)$ appears in  eq.~[\ref{eq:exprate}]). The density fluctuations are assumed to have grown by gravity out of small primeval irregularities, so the mass density contrast varies as $\delta\propto D(t)$, where $D(t)$ is the growing solution to the time evolution of $\delta$ in linear perturbation theory. The result of integrating equation~(\ref{eq:mass_con}) over a sphere of radius $R$ and applying Gauss's theorem is the wanted relation, 
\beq
	\bar v = -\int {d\Omega\over 4\pi }{\bf v}\cdot {\bf n}
	= {1\over 3}fH_oR\bar\delta .
\label{eq:vbar}
\eeq
Here ${\bf n}$ is the unit normal of the sphere, $\bar v$ is
the radial inward peculiar velocity averaged over the surface, 
and $\bar\delta =\delta M/M$ is the mass contrast
averaged within the sphere. The dimensionless factor $f$ is
\begin{equation}
	f = {\dot D\over D}{a\over\dot a} \simeq \Omega _m^{0.6}.
\label{eq:ffactor}
\end{equation}
The power law is a good approximation if $\Lambda =0$ or space
curvature vanishes.  

For a sphere centered on the Virgo Cluster, with us on the
surface, estimates of the mean radial flow through the sphere 
and the contrast $\delta _g = \delta M/M$ in galaxy counts within
the sphere are 
\beq
	\bar v=200\pm 25\hbox{ km~s}^{-1}, \qquad
	\bar\delta _g = 2.3\pm 0.7.
\label{eq:JTonry}
\eeq
The velocity is from a survey in progress 
by Tonry {\it et al.} (1998); it is consistent with
earlier measurements (eg. Faber \&\ Burstein 1988). The density
contrast in counts of IRAS galaxies within 
our distance from the Virgo cluster is $\bar\delta =1.4$ (Strauss
{\it et al.} 1992). IRAS galaxies are detected because they are
rich in gas and have high star formation rates, making them
prominent in the 60 to 100 micron range of the IRAS satellite
survey. IRAS galaxies avoid dense regions, likely because
collisions and the ram pressure of intracluster gas have stripped
the galaxies of the gas that fuels bursts of star
formation and high infrared luminosity; a
commonly used correction factor of 1.4 would bring the contrast
for optical galaxy counts to $\delta _g=2$. A 
preliminary analysis of the Optical Redshift Survey (Santiago
{\it et al.} 1996) by Strauss (1998) gives $\delta _g\sim 3$. 
With $\bar\delta =\bar\delta _g$,
equations~(\ref{eq:vbar}) to~(\ref{eq:JTonry}) give
\beq
	\Omega _m= 0.1{+0.1\atop -0.05}.
\label{eq:Omega_dyn}
\eeq
This is plotted as the right-hand point in Figure~1. 

There are three key assumptions. First, the analysis uses 
conventional gravity physics. An alternative, Milgrom's (1995)
modified Newtonian dynamics (MOND), has been quite durable in
applications to individual galaxies (de Blok \&\ McGaugh 1998).
An extension to the analysis of large-scale flows would be
interesting, but the focus here is the test of cosmology based on
the conventional gravity physics of general relativity 
theory. Second, the relation between peculiar velocities and the
mass distribution follows from the assumption that structure 
grew by gravity out of small primeval departures from
homogeneity. (Thus the boundary condition for
eq.~[\ref{eq:ffactor}] is $D(t)\rightarrow 0$ at
$z\gg 1$.) Most dynamical measures of $\Omega _m$ use this
assumption; the exception is relaxed systems that have forgotten
their initial conditions (as in the velocity dispersion measure
used by Marzke {\it et al.} 1995.) 
We have no viable alternative to the gravitational
instability picture for structure formation on large scales, but
it will be checked by consistency with all the other
cosmological tests, when they are better established. Third, and most contentious,
equation~(\ref{eq:Omega_dyn}) assumes the mass 
clusters with the galaxies. If the mass contrast were reduced to 
$\delta\sim 0.2\delta _g$ then the other numbers would
be consistent with $\Omega _m= 1$. The concept that the galaxy
distribution may be a biased measure of the mass distribution
has been influential, and rightly so; this important issue had to
be explored. But as discussed next I think it is also fair to say that there never was any evidence for what I would expect to be  the distinctive signature of biasing: void galaxies.  

If $\Omega _m=1$ we must decide where most of the mass is. It can't be in groups and clusters of galaxies: Figure~1 shows that analyses
similar to the above yield similar values of $\Omega _m$ in the
Local Group (Peebles 1996) and in clusters of galaxies 
(Carlberg {\it et al.} 1996). That leaves the voids, spaces
between the concentrations of giant galaxies. We 
know the voids are not undisturbed: absorption lines in
quasar spectra show that at redshift $z=3$ space was filled with
clouds of hydrogen. Thus voids would  
have to be regions where star or galaxy formation was suppressed. 
I find it hard to believe the suppression was so complete as to
leave nothing  observable; surely there would be irregulars or
dwarfs from almost failed seeds.  Searches in relatively nearby
voids, where galaxies are observable well into the  faint end of
the luminosity function, reveal no such population. Perhaps the  
gravitational growth of clustering swept the void galaxies into
the concentrations of normal ones, but in that case gravity would
have pulled the mass with the galaxies, suppressing biasing
(Tegmark \&\ Peebles 1998). Perhaps our picture for structure
formation needs tuning; that will be checked as the
cosmological tests improve. The straightforward reading is that biasing is not a strong factor; $\Omega _m$
is substantially less than unity. This is the basis for the grades
in line 1a in Table~2. The grades are  subject to negotiation, of
course; the discovery of a population of void galaxies would
make a big difference to me. 

The theory of the origin of the light elements requires baryon
density parameter  $\Omega _{\rm baryon} = 0.02/h^2\sim 0.04$ for
$h=0.7$.  It is not easy to reconcile the dynamical analyses 
with such a small value for $\Omega _m$. As I noted in the last section, the common assumption is that the mass of the universe is dominated by nonbaryonic dark matter. There certainly is
nothing unreasonable about the idea---Nature need not have
put most of the matter in a readily observable form---but the
cosmology certainly would be cleaner if we had a laboratory detection of this hypothetical mass component.

\subsection{Expansion Rate and Time}

Since we are considering what the \flm\ does and does not
predict we should note that the model allows solutions without
a Big Bang, that trace back through a bounce to contraction from 
arbitrarily low density. This requires $\Lambda >0$ and positive
space curvature, and, if the universe is going to contract by a
substantial factor before bouncing, very large space curvature and small matter density: the redshift at  
the bounce is $z_{\rm max}\sim |\Omega _R|/\Omega _m$. 
The bounce case is seldom mentioned, and I suspect rightly so, for
apart from the bizarre initial conditions the redshift
$z_{\rm max}$ required for light element production requires 
quite unacceptable density parameters. If this assessment is
valid we are left with \fl\ solutions that trace back to
infinite density, which is bizarre enough but maybe can be
finessed by inflation and resolved by better gravity physics. 

A \flm\ that expands from exceedingly high density predicts that
stellar evolution ages and radioactive decay ages are less than
the cosmological expansion time $t_o$. Numerical examples are 
\beqa
	H_ot_o &=& 2/3\ \hbox{ if }\ \Omega _m=1,\quad \Omega _R=0,
		\nonumber \\
	&=& 0.83\ \hbox{ if }\ \Omega _m=0.25,\quad \Omega _R=0.75,
	\label{eq:Hoto}\\
	&=& 1.01\ \hbox{ if }\ \Omega _m=0.25,\quad \Omega _R=0.
		\nonumber
\eeqa
The Hubble Space Telescope Key Project (Freedman
{\it et al.} 1998; Madore {\it et al.} 1998) reports
\beqa
	H_o&=&73\pm 6(\hbox{statistical})\nonumber \\
	&&\ \pm 8(\hbox{systematic}) 
	\hbox{ km s}^{-1}\hbox{ Mpc}^{-1}.
\eeqa
The systematic error includes length scale calibrations common to
most measurements of $H_o$. A recent survey of evolution ages of
the oldest globular cluster stars yields $11.5\pm 1.3$~Gyr. 
(Chaboyer {\it et al.} 1998). We have to
add the time for expansion from very high
redshift to the onset of star formation; a commonly used nominal
value is 1~Gyr. If the universe is 14~Gyr old this would put the
onset of star formation at $z\sim 5$ in the Einstein-de~Sitter
model, $z\sim 6$ if $\Omega _m= 0.25$ and 
$\Omega _\Lambda = 0.75$. Since star forming galaxies are
observed in abundance at $z\sim 3$ (Pettini {\it et al.}  1998
and references therein) this is conservative. These numbers give 
\beq
	H_ot_o = 0.93\pm 0.16,
\eeq
where the standard deviations have been added in quadrature. 

The result agrees with the low density models in
equation~(\ref{eq:Hoto}). The Einstein-de~Sitter case is
off by 1.6 standard deviations, not a serious discrepancy. It
could be worse: Pont {\it et al.} (1998) put the minimum stellar
evolution age at 13~Gyr. With 1~Gyr for star formation this would
make the Einstein-de~Sitter model $>2.6\sigma$ off. It could go
the other way: an analysis of the distance scale implied by the
geometry of the multiply lensed system PG~1115+090 by Keeton \&\
Kochanek (1997) puts the Hubble parameter at $h=0.51\pm 0.14$. At 
$t_o = 14$~Gyr this says $H_ot_o=0.73\pm 0.20$, nearly
centered on the Einstein-de~Sitter value. An elegant argument
based on the globular cluster distance to the Coma Cluster of
galaxies leads to a similar conclusion (Baum 1998). 
Most estimates of $H_o$ are larger, however, and the correction
to $t_o$ for the time to abundant star formation is conservative,
so in line 1b of Table~2 I give the Einstein-de~Sitter model a modest demerit for its expansion time. 

The low density cases pass the time-scale constraint at the
accuracy of the present measurements. Since a satisfactory and
it is to be hoped feasible measurement would distinguish between the $\Omega _m\sim 0.25$ open and flat cases I lower their 
grades from this test to $\surd$?. 

\subsection{Probes of Spacetime Geometry}

If spacetime is close to homogeneous and isotropic and described
by a single line element then the geometry is represented by two
functions of  redshift:  $r(z)$ fixes the angle subtended by
an object of given linear size at redshift $z$, and $dt/dz$ fixes proper world time as a function of redshift. The latter determines
$H_ot_o$, the former the $z$-$m$ relation. In the 
measurements of the $z$-$m$ relation by Perlmutter {\it et al.}
(1998a,b) and Reiss {\it et al.} (1998) the cosmologically
flat \flm\ with $\Omega _m= 0.25$ and $\Omega _\Lambda = 0.75$ is
within one standard deviation, the open model 
with $\Lambda =0$ and $\Omega _m= 0.25$ is about $3\sigma$ 
off, and the Einstein-de~Sitter model is some $7\sigma$ off. I explained in \S 2.2 why I suspect the
case against the open low density model is serious but maybe
premature: we should await further consideration of these new measurements by the authors and the community. The 
Einstein-de~Sitter model would require a more substantial 
reconsideration, so it gets a more serious demerit in line 1c in Table~2. 

Both functions, $r(z)$ and $dt/dz$, enter galaxy counts and the
rate of lensing  of quasars by the gravitational deflection of
the masses in foreground galaxies.  
The importance of the latter was demonstrated by  
Fukugita, Futamase, \&\ Kasai (1990) and Turner (1990). The
analysis by Falco, Kochanek, \&\ Mu\~ noz (1998) indicates 
that, in a cosmologically flat model, $\Omega _m\gsim 0.38$ at
$2\sigma$. An open low density model does better: $\Omega _m=0.25$ is at the $2\sigma$ 
contour. 

This constraint from lensing depends on the galaxy mass function. 
The predicted peak of the lensing rate at angular separation
$\theta\sim 1$ arc~sec is dominated by the high surface density 
branch of early-type galaxies at luminosities $L\sim L_\ast$
(where the galaxy mass function is approximated as
$dn/dL\propto L^\alpha e^{-L/L_\ast}$, with $\alpha\sim -1$).
The number density of these objects is not well known, 
because it is difficult to separate counts of early-type galaxies
in the high surface density branch from a low density
branch that is likely to be irrelevant for lensing (Kormendy 1987). Masataka Fukugita and I  have been unable to find a reliable way around this ambiguity using available surveys. 


If further tests of the lensing and redshift-magnitude
constraints confirmed the apparent inconsistency in entries 1c
and 1d the lesson could be that the cosmological constant is
dynamical, rolling to zero, as Ratra \&\ Quillen (1992) point
out.  

I keep a line in Table~2 for counts because galaxies are
observed at redshifts greater than unity, where the predicted
counts are quite sensitive to the cosmological parameters. The
counts are quite sensitive to galaxy evolution, too, but people
may learn how to deal with that as the understanding of galaxy
evolution improves.

\subsection{Fluctuations in the Distributions of Mass and the CBR}

As noted in \S\S 2.3 and 3.1, structure
formation on the scale of  galaxies and larger is thought to have
been dominated by the gravitational growth of small departures 
from homogeneity present in the very early universe. The nature
of the initial conditions is open because we do not have an
established theory of what the universe was doing before it was
expanding. We do have a consistency condition, that a single set
of initial values must match many observational constraints.
I discuss here second moments of the large-scale fluctuations in
the distributions of galaxies and the thermal cosmic background
radiation (the CBR).   

It is sensible to try the simplest prescription for initial
conditions first. Most widely discussed is the adiabatic cold
dark matter (ACDM) model Joe Silk mentions in his  introduction.
In the simplest case the universe is  Einstein-de~Sitter and the
density  fluctuations are scale-invariant (the density  contrast
$\delta\rho /\rho$ appearing on the Hubble length is independent
of  time). This case tends to underpredict large-scale density
fluctuations; the problem is remedied by lowering $H_o$ or 
$\Omega _m$ (Blumenthal, Dekel, \& Primack 1988; Efstathiou,  Sutherland, \& Maddox 1990).\footnote{Lowering $H_o$ or $\Omega _m$ lowers the expansion rate at the epoch of equality of mass densities in matter and radiation, and the larger expansion time when the universe is dominated by the pressure of the CBR increases the clustering length.} The wanted value of $H_o$ is below most estimates of this parameter, so the more commonly accepted interpretation is that $\Omega _m$ is less than unity. This leads to the grade in line 2a. It depends on the model for structure formation, of course.  

Examples of second moments of the galaxy space distribution and the angular distribution of the CBR are shown in Figures~2 and~3.
The power spectrum of the space distribution is
\beq
        P(k) = \int d^3r\,\xi (r)
        e^{i{\bf k}\cdot {\bf r}}, 
\eeq
where the dimensionless galaxy two-point correlation function is
\beq
	\xi (r) = \langle n({\bf r} + {\bf y})n({\bf y})\rangle
	/\langle n\rangle ^2 - 1,
\eeq
for the smoothed galaxy number density $n({\bf r})$. The data in
Figure~2 are from  the IRAS PSC-z (point source catalog) redshift
survey (Saunders {\it et al.}  1998) of the far
infrared-luminous galaxies mentioned in \S 3.1. Since infrared
radiation is not strongly affected by dust this promises to be an 
excellent probe of the large-scale galaxy distribution. 

The expansion in spherical harmonics of the CBR temperature as a function of direction in the sky is
\beq
	T(\theta ,\phi) = \sum a_l^m Y_l^m(\theta ,\phi ).
\eeq
Figure~3 shows second moments of the expansion, defined as
\beq
	T_l = \left[ {l(l + 1)\over 2\pi } \right] ^{1/2}
	\langle |a_l^m|^2\rangle ^{1/2}.
\label{eq:Tl}
\eeq
In the approximation of the sum over $l$ as an integral 
the variance of the CBR temperature per 
logarithmic interval of $l$ is $(T_l)^2$. The $T_l$ data in Figure~3 are from the survey of the measurements by Tegmark (1998a). 

The solid curves in figures~2 and~3 are the prediction (Tegmark
1998a,b) of an ACDM model with a scale-invariant primeval mass
fluctuation spectrum and the parameters 
\beqa
	&&\Omega _m= 0.8, \quad\Omega _\Lambda = -0.1,
	\quad\Omega _R = 0.3,\nonumber \\
	&&\Omega _{\rm baryon} = 0.1, \qquad h = 0.5.
\label{eq:Max}
\eeqa
It is impressive to see how well this model fits the two sets of measurements. But at the present accuracy of the measurements there is at least one other viable model, shown as the dashed curves. It assumes the same  dynamical actors as in ACDM---cold
dark matter, baryons, the CBR, and three families of massless
neutrinos---but the isocurvature initial condition is that the
primeval mass density and the entropy per baryon are homogeneous,  
and homogeneity is broken by an inhomogeneous primeval
distribution of the CDM. A simple model for the spectrum of
primeval CDM fluctuations is $P(k)\propto k^m$. A rough fit to the
measurements has parameters
\beqa
	&&\Omega _m= 0.2, \quad\Omega _\Lambda = 0.8, 
	\quad\Omega _R = 0,\nonumber \\
	&&\Omega _{\rm baryon} = 0.03, \quad
	 h = 0.7,\quad m = -1.8.\label{eq:parameters}
\eeqa
Further details and a pedigree within the inflation picture are in Peebles (1999a, b). The solid curve fits the CBR anisotropy measurements better, but it is based on a much more careful search of parameters to fit the data. A bend in $P(k)$ would do wonders for the dashed curve. Hu (1998) gives another example of how the prediction of the CBR angular fluction spectrum depends on the details of the structure formation model. 

As mentioned in \S 2.3, the point of this discussion is that reading the values of the cosmological parameters from the CBR anisotropy measurements in Figure~3 is a dangerous operation because it depends on the theory for structure formation as well as the \fl\ model. This applies to other entries in category~2 in Table~2 (and to line 1a: a satisfactory quantitative understanding of galaxy formation would include an understanding of the relation between the distributions of galaxies and mass). 

Our knowledge of $P(k)$ and $T_l$ will be
considerably improved by work in progress. Redshift surveys to
probe $P(k)$ and the large-scale mass distribution include the
Century Survey, the Two Degree Field Survey (2dF), and the Sloan 
Digital Sky Survey (SDSS); precision measurements of the CBR
include BOOMERANG, MAP, PLANCK, and other ground, balloon, and satellite projects (Geller {\it et al.} 1997; Page 1997; Nordberg \&\ Smoot 1998; Eisenstein, Hu, \&\  Tegmark 1998; and references therein). If one of the structure formation models now under discussion fits all the bumps and wiggles in the measured spectra it will inspire confidence.

In the Einstein-de Sitter case a scale-invariant ACDM model normalized by the assumption that galaxies trace mass gives quite a good fit to the CBR angular fluctuation spectrum $T_l$; on this score it would merit a pass in line 2b. But the assumptions that galaxies trace mass and that $\Omega _m=1$ imply quite unacceptable peculiar velocites. The situation is different from line 1a, where the issue is whether $\Omega _m=1$ can be saved by the postulate that galaxies do not trace mass. Thus I think it is fair to give the Einstein-de~Sitter case separate demerits in lines 1a and 2b, but with a question mark for the latter because it depends on the model for structure formation.

Several authors have concluded that the low density flat ACDM model (with $\Lambda > 0$) is a better fit to the $T_l$ measurements than is the low density open case (Gawiser \&\ Silk, 1998; Tegmark 1998a). Others note that other treatments of the still quite new measurements can lead to the opposite conclusion (G\'orski {\it et al.} 1998; Ratra 1998). Since the former approach seems to treat the measurements in the more literal way the flat case gets the higher grade in line 2b.

\subsection{The Evolution of Clusters of Galaxies}

Bahcall and colleagues (Bahcall \&\ Fan 1998 and references
therein) have emphasized the importance of the time evolution of
the number of clusters of galaxies as a probe of the cosmology.
The condition that a CDM model fit the present cluster number density is (White, Efstathiou, \&\ Frenk 1993; Bahcall \&\ Fan 1998) 
\beq
	\sigma _8 = 0.53\Omega _m^{-0.53},
\label{eq:sigma8}
\eeq
where $\sigma _8=\delta M/M$ is the rms contrast in the mass
found within a randomly placed sphere of radius $8h^{-1}$~Mpc. 
Since the rms fluctuation of galaxy counts is close to unity on
this scale, $\sigma _8(g)\simeq 1$, equation~(\ref{eq:sigma8}) says galaxies trace mass if $\Omega _m\sim 0.3$, while biasing has to be substantial if
$\Omega _m=1$. I have already indicated why I am skeptical of the
latter. More important, Bahcall \&\ Fan~(1998) demonstrate that with the Einstein-de~Sitter parameters the ACDM model normalized to fit the present cluster number density quite underpredicts the abundance of clusters at $z\gsim 0.5$. 

This result assumes Gaussian initial density fluctuations. If
$\Omega _m=1$  the present mass fraction in clusters is small, so
the normalization is to a steeply falling part of the Gaussian. 
The time evolution of the rms density fluctuation consequently
causes a large evolution in the predicted number of clusters. The 
Gaussian case is simple and natural to consider first, and it
follows from simple models for inflation, but there are other 
possibilities. In the ICDM  model (\S 3.4) the CDM could be a
massive field squeezed from its ground level during inflation, 
in which cases the primeval CDM mass distribution is  
$\rho ({\bf r})=m^2\phi ({\bf r})^2/2$,
where $\phi$ is a random Gaussian process with zero mean. In this model the mass fluctuation distribution is much less steep than a 
Gaussian, the cluster abundance accordingly is a less sensitive
function of the rms mass  fluctuation, and the Einstein-de~Sitter
model predicts acceptable evolution of the cluster mass function 
(Peebles 1999b). It is not clear whether the constraint from the  skewness of the galaxy count distribution (Gazta\~naga \& Fosalba 1998) allows the primeval mass fluctuations to be non-Gaussian enough for acceptable cluster evolution in the Einstein-de~Sitter case. 

The evolution of structure is a key probe of cosmology, and Bahcall and colleagues have demonstrated that the rich clusters of galaxies offer a particularly sensitive measure. But I am inclined to keep the question marks on the grades in line 2c until we can be more sure of the nature of the initial conditions for structure formation. 

\subsection{Cluster Baryon and Dark Matter Masses}

This important probe was pioneered by White {\it et al.} (1993). In the survey by White \& Fabian (1995) the ratio of the mass in X-ray emitting gas to the gravitational mass in rich clusters of galaxies is
\beq
	(M_{\rm HII}/M_{\rm grav})_{\rm cl}=(0.056\pm 0.014)h^{-3/2}.
\eeq
Myers et al. (1997) find from the measurement of the
Sunyaev-Zeldovich effect in three clusters
\beq
	(M_{\rm HII}/M_{\rm grav})_{\rm cl}=(0.061\pm 0.011)h^{-1}.
\eeq
In their contributions to this discussion Silk and Turner explain why the consensus value of the density
parameter in baryons to account for light element abundances is 
$\Omega _{\rm baryon} = 0.02/h^2$. If clusters are fair samples
of baryon and total masses then $\Omega _{\rm baryon}/\Omega _m$
is the same as $(M_{\rm baryon}/M_{\rm grav})_{\rm cl}$. If most
of the cluster baryons are in the plasma we get from these mass
ratios $\Omega _m= (0.36\pm 0.09)h^{-1/2}$
and $\Omega _m= (0.33\pm 0.06)h^{-1}$. The correction for baryons in stars decreases $\Omega _m$. Energy injected by winds from supernovae in cluster galaxies would tend to lower the plasma mass; a correction for this effect would further lower $\Omega _m$ (Metzler \&\ Evrard 1998). At $h>0.5$ this measure of $\Omega _m$ is well below  Einstein-de~Sitter.  

On the other hand, if baryons settled to cluster centers, increasing the local ratio of baryon to total mass, it would bias this measure of $\Omega _m$ low. It is not hard to make up a story for how this might have happened. Imagine that before there were clusters there were gas clouds dense enough that the baryons dissipatively settled, leaving dark matter halos. We have to postulate the clouds were small enough that the radiation from this dissipative settling is not objectionably hard, and we have to postulate that feedback from star formation prevented catastrophic collapse of the baryons. Now imagine many of these systems fall together to form a proto-cluster. Numerical N-body simulations of merging show that the dense parts of the substructure tend to settle relative to less dense parts, producing the wanted segregation of baryons from the dark matter. Numerical simulations of cluster formation fail to show any evidence of this story; I do not know whether that is because it is only a story or possibly because it is hard to explore all scenarios in numerical simulations. 

One does hears mention of the possibility of an inhomogeneous primeval entropy per baryon, but with little enthusiasm. 

As indicated in line 2d this constraint on $\Omega _m$ so far has proved difficult to finesse.

\subsection{Pure Thought}

There are three issues to consider: coincidences, inflation, and our taste as to how the world might best end. 

If $\Omega _m\sim 0.25$ we flourish at a special epoch, just as the universe is making the transition from matter-dominated to
$\Lambda$- or curvature-dominated expansion  within the \flm .
Maybe this is pure chance. Maybe it is an effect of selection: 
perhaps galaxies as homes for observers would not have existed if
$\Omega _m$ were very different from unity (Martel, Shapiro \&\ Weinberg 1998 and references therein). Maybe there is no coincidence: perhaps $\Omega _m$ really is close to unity. Most of us consider the last the most reasonable possibility. But the observational entries in Table~2 show that
if $\Omega _m=1$ then Nature has presented us with a considerable
set of consistently misleading clues. The much more likely reading of the evidence is that, within the \flm , $\Omega _m\simeq 0.25$. We should pay attention to arguments from aesthetics; the history of physical science has many  examples of the success of ideas driven by logic and elegance. But there are lots of examples of surprises, too. The evidence that $\Omega _m$ is significantly less than unity is a surprise. I enter it as a demerit for the \flm , but not a serious one: surprises happen. 

The conventional inflation picture accounts for the near
homogeneity of the observable universe by the postulate that an 
epoch of near exponential expansion driven by a large effective
cosmological constant stretched all length scales in the primeval  chaos to unobservably large values, making the universe we can see close to uniform. The same process would have made the
radius of curvature of space very large. Thus in their book, 
{\it The Early Universe}, Kolb \&\ Turner (1988)
emphasize that a sensible inflation theory requires negligibly
small space  curvature: $\Omega _m$ may be less than unity, but
if it is a cosmological constant makes 
$\Omega _m+\Omega _\Lambda$ equal to unity. The argument is
sensible but 
model-dependent. Gott (1982) pioneered a variant of inflation
that produces a near homogeneous \flm\ with open space sections.
Ratra \&\ Peebles (1994) revived the concept; details of the
history and application are in G\'orski {\it et  al.} (1998).
Most proponents of inflation I have talked to share the
preference for $\Omega _m+\Omega _\Lambda = 1$ but agree that
they could learn to live with the open version if that is what
the observations required. The flat low density case does get
the higher grades in Table~2, from the $z$-$m$ relation and
the CBR anisotropy, but my impression is that both results are too preliminary to support a decision on open versus flat space sections.

The values of the cosmological parameters tell us how the world ends according to the \flm , whether it is collapse back to a Big Crunch or expansion into the
indefinitely remote future. But why should we pay attention to
an extrapolation into the remote future of a theory
we can be pretty sure is at best only an approximation to reality? For example, suppose improved tests showed that $\Omega _m=0.2$, that the dimensionless measure $\Omega _\Lambda$ of Einstein's cosmological constant is quite small, $|\Omega _\Lambda |\ll 1$, and that space curvature correspondingly is negative. The straightforward 
interpretation would be that our universe is going to expand
forever more, but it need not follow. If $\Lambda$ were 
constant and less than zero, then no matter how small 
$|\Omega _\Lambda |$ the \flm\ would predict that the expansion will eventually stop and the universe will contract back to a Big Crunch.\footnote{The example is contrived but not entirely frivolous. Standard and successful particle theory includes a cosmological constant, in
the form of the energy density of the vacuum, but quite fails to
explain why its value is in the observationally acceptable range.
Until we have a deeper theory that deals with this I don't see
how we can exclude the idea that it has or ends up with an exceedingly small negative value.} This may be of some comfort if the Big Crunch is more to your taste. To my taste the main lesson is that we should stop all this talk about how the world ends until we can think of some scientific meaning to attach to the answer. 

\section{Concluding Remarks}

It is impressive to see how well the \flm\ fits the full range
of observations summarized in Table~2. We have to to bear in mind
that many of the measurements still are open to
discussion, however, and that the entries in category~2 depend on a model for structure formation that also has to be tested. Thus there is a large number of question marks (even though I believe I have been an easy grader). Perhaps the best lesson one might draw from the length of the discussion of Table~2 in \S 3 is that we theorists ought to resist the temptation to draw large conclusions from the latest observational reports; these are extraordinarily difficult measurements that we best praise by respectful cautious consideration. 

I think we should also bear in mind that substantial parts of the left-hand column of Table~1 were formulated a full seven decades ago, and that much of the rest was driven by observational advances. That is, although we have many elegant new theoretical ideas in cosmology, we have little evidence in hand on which Nature has chosen. 

The right-hand column of Table~1, that represents the observational constraints, is considerably longer than it would have been in a list made ten years ago, and ten years ago there would have been a lot more question marks in Table~2. We can be sure work in  progress will produce a
considerably tighter network of cosmological tests ten years from now. I see no reason to think the results will fail to support the \flm , but that will be revealed in the fullness of time and a lot of hard work.

\acknowledgements

I am grateful to David Hogg, Wayne Hu, and Max Tegmark for stimulating discussions. This work was supported in part at the Princeton Institute for Advanced Study by the Alfred P. Sloan Foundation.

\begin{figure}
\centerline{\psfig{file=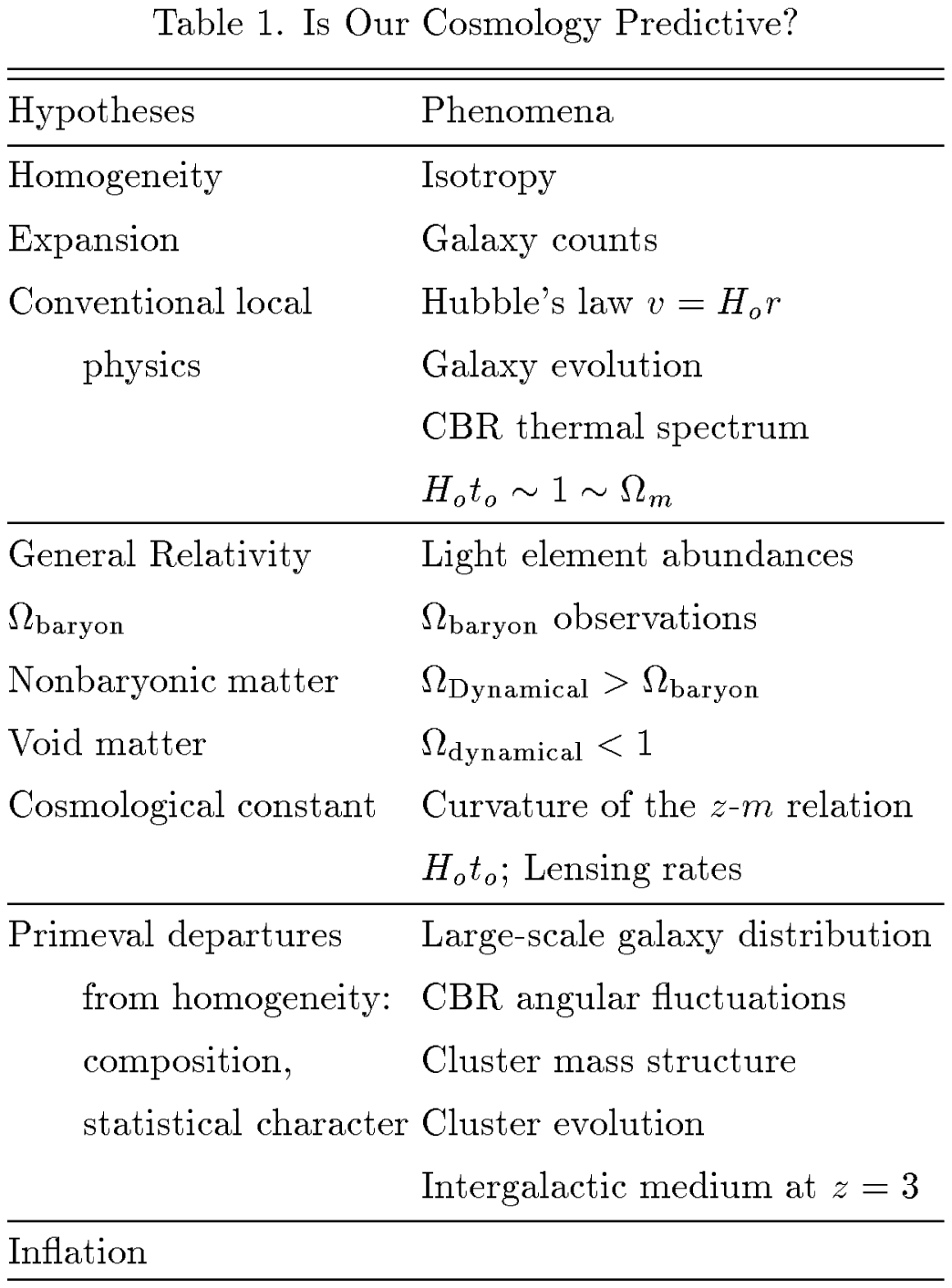,width=3.2truein,clip=}}
\end{figure}

\begin{figure}
\centerline{\psfig{file=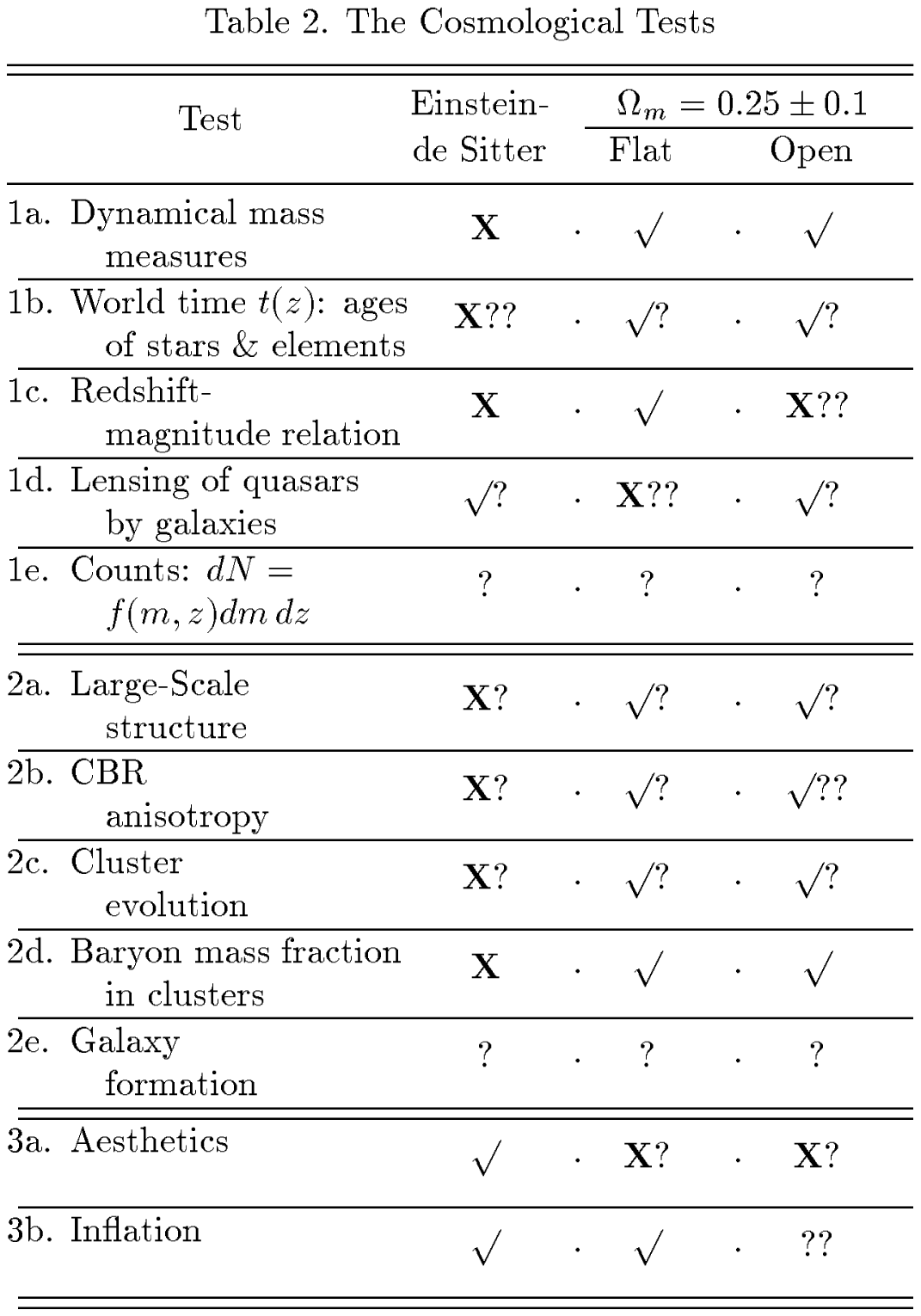,width=3.2truein,clip=}}
\end{figure}

\begin{figure}
\centerline{\psfig{file=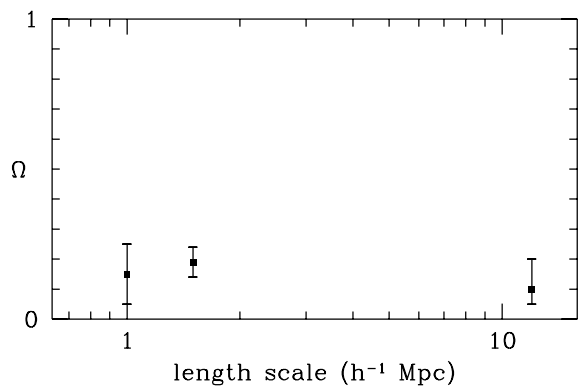,width=3.truein,clip=}}
\caption{Dynamical estimates of the mean mass density
from galaxy velocities in and near the Local Group (left point),
in rich clusters of galaxies, and from the flow of galaxies
toward the Local Supercluster (right-hand point).}
\end{figure}

\begin{figure}
\centerline{\psfig{file=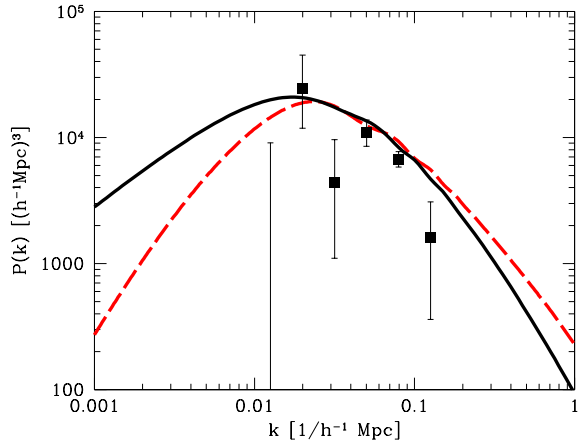,width=3.truein,clip=}}
\caption{Mass fluctuation spectrum extrapolated to
the present in linear perturbation theory for the ACDM model in
equation~(\ref{eq:Max}) (solid line, from Tegmark 1998b) and the
ICDM model in equation~(\ref{eq:parameters}) (dashed line). The
galaxy fluctuation spectrum is from the PSC-z collaboration
(Saunders {\it et al.} 1998).} 
\end{figure}

\begin{figure}
\centerline{\psfig{file=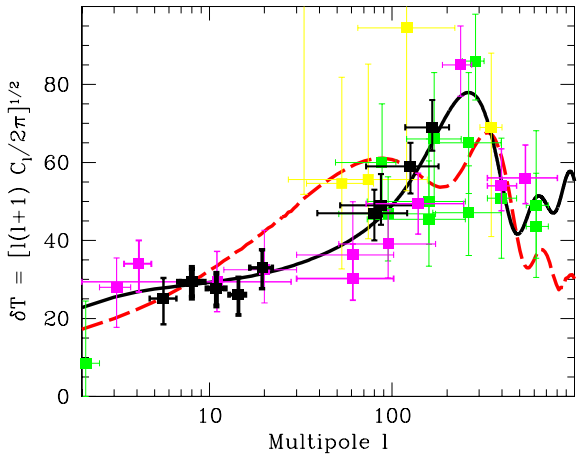,width=3.truein,clip=}}
\caption{Spectrum of angular fluctuations of the
CBR. The data are from the compilation by Tegmark (1998a).
The ACDM model prediction plotted as the solid line assumes the
parameters in equation~(\ref{eq:Max}) (Tegmark 1998a). The ICDM
model prediction plotted as the dashed line assumes the
parameters in equation~(\ref{eq:parameters}).}
\end{figure}

\end{document}